\documentclass[conference]{IEEEtran}
\IEEEoverridecommandlockouts
\newtheorem{definition}{Definition}

\usepackage{graphicx}

\usepackage{color}
\usepackage{hyperref}
\makeatletter
\newcommand{\newlineauthors}{%
  \end{@IEEEauthorhalign}\hfill\mbox{}\par
  \mbox{}\hfill\begin{@IEEEauthorhalign}
}
\makeatother
\AtBeginDocument{%
  \providecommand\BibTeX{{%
    \normalfont B\kern-0.5em{\scshape i\kern-0.25em b}\kern-0.8em\TeX}}}

\author{\IEEEauthorblockN{Kaneez Fizza}
\IEEEauthorblockA{\textit{Swinburne University of Technology}\\
Melbourne, Australia, \\
kfizza@swin.edu.au}
\and
\IEEEauthorblockN{Prem Prakash Jayaraman}
\IEEEauthorblockA{\textit{Swinburne University of Technology}\\
Melbourne, Australia, \\
pjayaraman@swin.edu.au}
\and
\IEEEauthorblockN{Abhik Banerjee}
\IEEEauthorblockA{\textit{Swinburne University of Technology}\\
Melbourne, Australia,  \\
abanerjee@swin.edu.au}

\newlineauthors
\IEEEauthorblockN{Dimitrios Georgakopoulos}
\IEEEauthorblockA{\textit{Swinburne University of Technology}\\
Melbourne, Australia,  \\
dgeorgakopoulos@swin.edu.au}
\and
\IEEEauthorblockN{Rajiv Ranjan}
\IEEEauthorblockA{\textit{Newcastle University}\\
England, United Kingdom \\
rranjans@ncl.ac.uk}
}

\begin{document}
\title{Evaluating Sensor Data Quality in  Internet of Things Smart Agriculture Applications}
\maketitle
\begin{abstract}
The unprecedented growth of Internet of Things (IoT) and its applications in areas such as Smart Agriculture 
compels the need to devise newer ways for evaluating the quality of such applications.
While existing models for application quality focus on the quality experienced by the end-user (captured using likert scale), IoT applications have minimal human involvement and rely on machine to machine communication and analytics to drive decision via actuations. 
In this paper, we first present a conceptual framework for the evaluation of IoT application quality. Subsequently, we propose, develop and validate via empirical evaluations a novel model for evaluating sensor data quality that is a key component in assessing IoT 
application quality.
We present an implementation of the sensor data quality model and demonstrate how the IoT sensor data quality can be integrated with a Smart Agriculture application. Results of experimental evaluations conducted using data from a real-world testbed concludes the paper.

\end{abstract}

\maketitle



\section{Introduction}
\begin{figure}[htbp]
  \centering
  \includegraphics[width=85mm, scale=0.5]{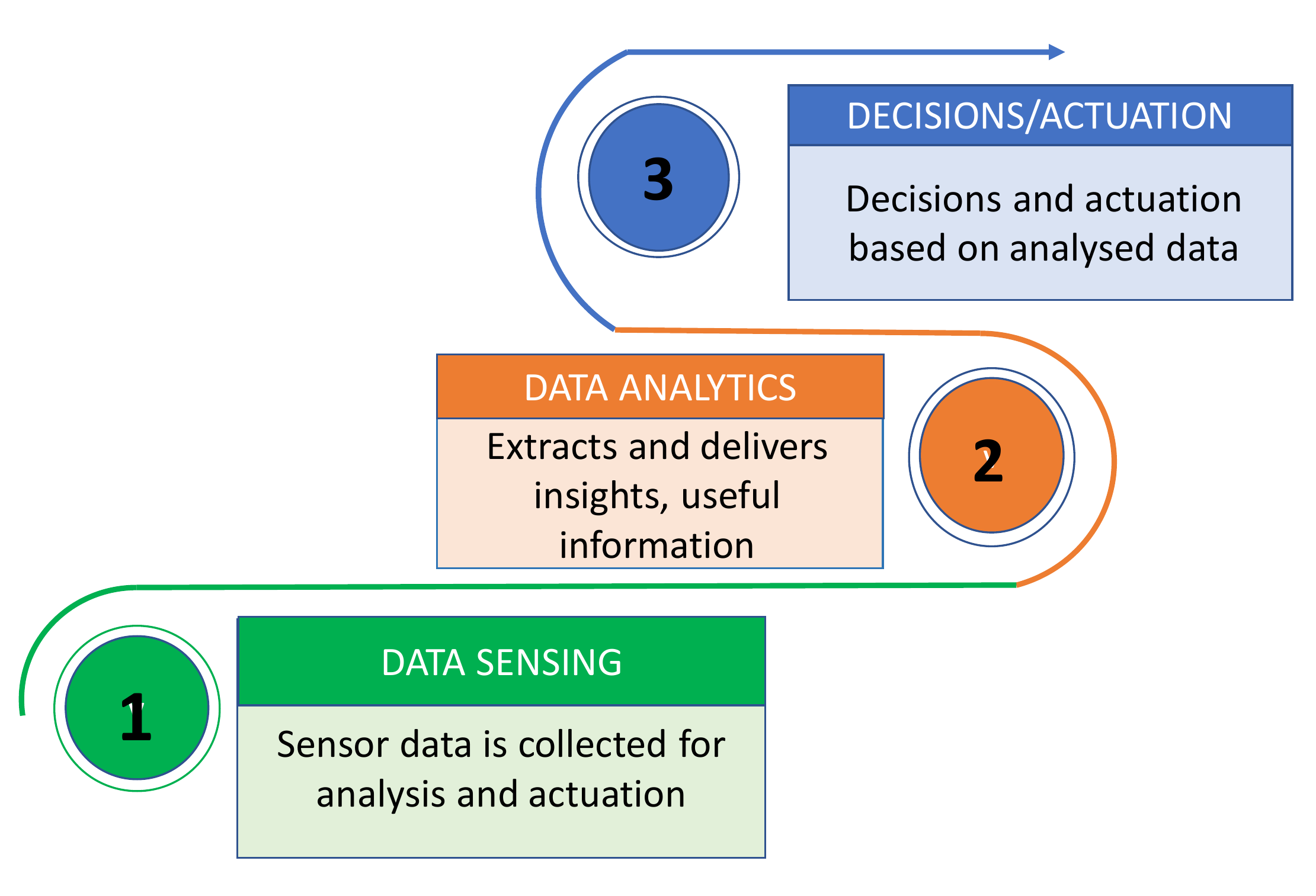}
  \caption{Stages in IoT application lifecycle}
 \label{fig:app_stage}
\end{figure}

The unprecedented growth of IoT and its applications in areas such as Smart Agriculture has put forward the need to devise novel ways of evaluating IoT application quality. Existing ways of evaluating application quality (e.g. using likert scale) cannot be directly adopted in IoT due to the autonomic nature of IoT.  
IoT applications collect data (e.g. about soil moisture and solar radiation), extract insightful information (e.g. determine growth of the plant) based on which they make decisions and actuation (e.g. irrigate the field). This lifecycle is depicted in Fig. \ref{fig:app_stage} and each stage impacts the quality of IoT application (e.g. was the amount of irrigation appropriate). Given the limited human involvement in the IoT application lifecycle and lack of human feedback, devising objective ways of evaluating the quality of IoT application is a grand challenge. In the context of this paper, quality of an IoT application is the ability of the IoT application to meet its key performance indicators (KPIs). For example, IoT application's ability to irrigate the farm at the right time, with the right amount of water based on data obtained from IoT sensors deployed in the farm. If the IoT sensor data are incorrect or misleading, derived decisions (via actuation) are likely flawed, and thus impacts the ability of the IoT application to meet its KPIs. Evaluating application quality is important as this allows IoT application designers to develop IoT applications that are resilient, robust and make actuation decision based on the underlying IoT sensor data quality.

There are multiple reasons that impact sensor data such as outlier \cite{yu2017recursive}. These faults adversely impact the quality of the data. In existing literature \cite{maurer2006activity}, \cite{nesa2017iot}, the quality of sensor data is calculated based on some ground truth (GT). This GT can come from multiple sources, from another sensor (called reference sensor), existing historical data or an alternate data source. However, the quality of the reference sensor itself is not guaranteed, the historical data may not always be available and the additional cost constrains an alternate data source \cite{ha2020sensing}. In the existing literature 
\cite{banerjee2017iot}, \cite{tu2018data}, 
authors have identified the importance of IoT sensor data as a key factor that influence the quality of an IoT application. They have characterized IoT data in terms of accuracy, completeness, suitability, usefulness etc. However, these works do not provide a single value for a composite quantitative measure of sensor data quality.


In this paper we proposes a model to assess the quality of IoT data collected from sensors during the sensing stage of an IoT application (Fig. \ref{fig:app_stage}). The proposed model provides a composite measure of sensor data quality, which can be used by the application designer to design quality-aware IoT applications (i.e. cope with low-quality sensor data in the application's decision making process i.e. actuation). We present a proof-of-concept implementation of the proposed model and demonstrate its efficacy via a smart agriculture cold chain management usecase \cite{jayaraman2015yourself}, \cite{jayaraman2015addressing}.




To the best of our knowledge, this is the first work that proposes and evaluates a model to calculate a single value for the quality measure of an IoT application based on the objective measurement collected from IoT sensors. The specific contributions of this paper are as follows:
\begin{itemize}
    \item We identify and define the factors which impact data quality (we call these factors quality metrics) during the sensing stage of an IoT application.
    \item A quality model for computing and evaluating the quality of IoT application based on sensor data and application context.
    \item A case study of smart agriculture cold chain monitoring IoT application to demonstrate the applicability of our proposed quality model.
\end{itemize}

This paper is organized as follows: While Section II summarizes related work, we provide an overview of Quality Aware IoT Application Development in section III. Section IV describes 
our proposed quality model for evaluating sensor data quality. In section V, we perform a case study and provide an evaluation of our proposed quality model. Section VI provides a conclusion and future research directions of this work.

\section{Related Work}
There is a growing interest in evaluating quality of sensor data due to  unprecedented growth in Internet of Things (IoT). In \cite{hossain2011modeling}, 
authors proposed a model that computes the quality of sensor data to help a user or a system in making an informed decision and improve the automated monitoring process. 
The proposed model  provides evaluation of data quality based on three metrics i.e. certainty, accuracy and timeliness. However, they use data from multiple sensors to arrive at the target information. \par

The authors in \cite{kuemper2018valid} proposed a framework which can generate a quality vector. This quality vector  consists of five quality metrics i.e. completeness, timeliness, plausibility, artificiality and concordance. The framework can be attached as a component to any IoT platform to provide pluasibility analysis for heterogeneous data sources In IoT. However, the framework is based on variety of data (can be streaming or historic) and from multiple data sources.

Our proposed model evaluates sensor data quality from the perspective of the application designer. The model computes and evaluate sensor data quality based on the context information and the factors which impact data quality. Moreover, our model do not require an alternate data source to measure the quality of data from the target sensor.

\section{Quality Aware IoT Application Development}
Existing methods of quality evaluation of IoT applications focus only end-user quality of experience (QoE). However, for IoT applications, it is often difficult for an application designer to assess the application quality using end-user QoE alone. As the IoT application stages involve computation in physically separate locations (such as multiple farms across the agriculture supply chain), it is difficult for an application designer to determine the reasons for not meeting the desired KPIs, since the quality of application is influenced by quality of data/information received at each stage of the IoT application life cycle (depicted in Fig. \ref{fig:app_stage}). Incorporating the notion of application quality during application design allows application designer to develop IoT application that can adapt to quality variations, both at the time of application design as well as during operation. For example, an application that provides weather-based alerts to farmers, may switch to a different source if the quality of remotely deployed farm weather station sensor is found to be inappropriate (producing low quality data).  

\begin{figure}[htbp]
  \centering
  \includegraphics[width=88 mm, scale=0.35]{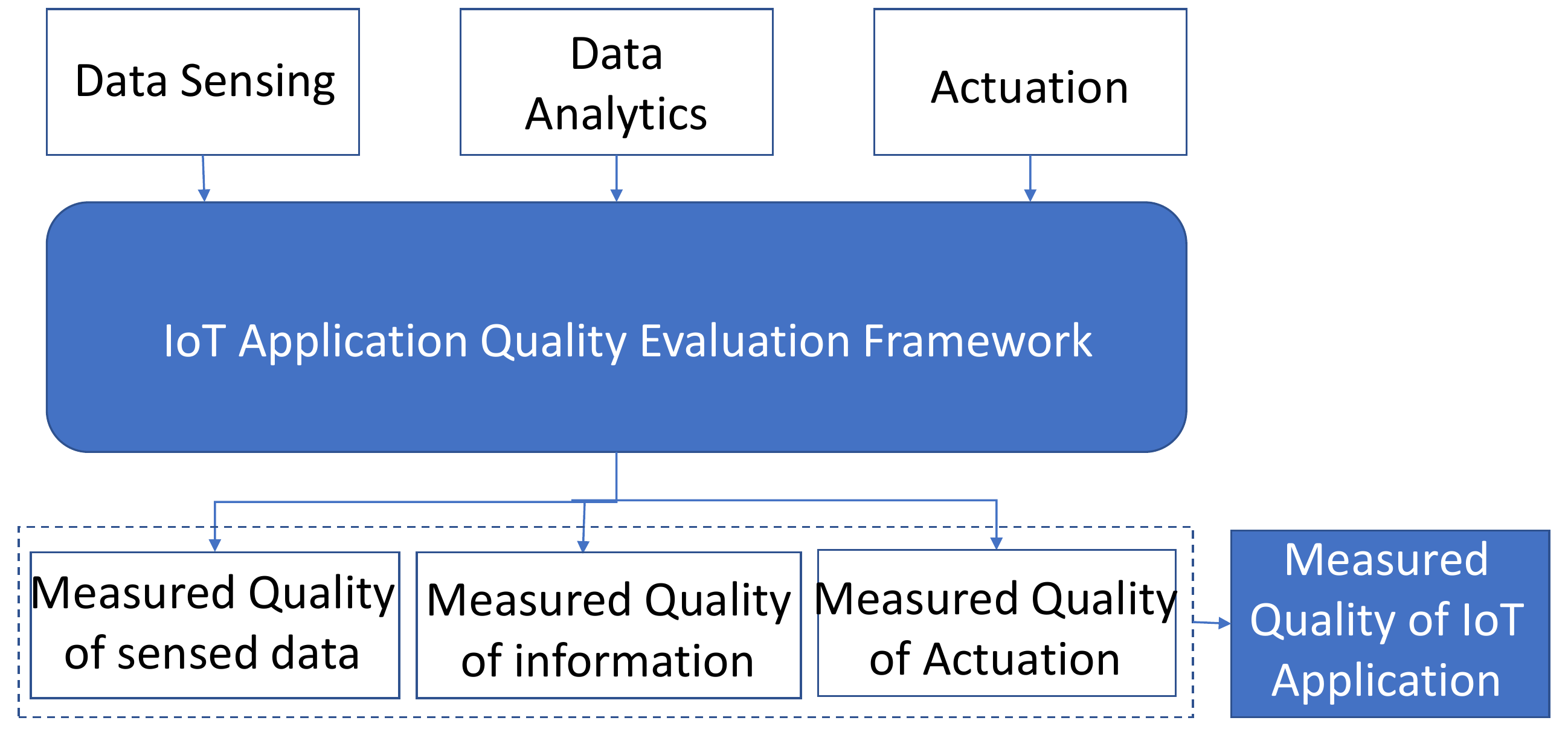}
  \caption{Framework for Evaluating IoT Application Quality}
 \label{fig:framework}
\end{figure}

 Fig. \ref{fig:framework} depicts the framework for evaluating quality of IoT application. 
The framework takes input at each stage of the application and produce quality as output at every stage. The overall quality of application is measured as an aggregated quality value from all the stages of the application. The key benefit of the proposed framework is that all the quality measurements are made available to the IoT application through a single programmatic interface. We propose our conceptual framework that monitors the quality at each stage of the IoT application, and provides these as inputs for the IoT application to consider during its decision making process. In this paper, we only focus on the computation of sensor data quality presented in subsequent sections.


\section{A Quality Model for Evaluating Sensor Data}
Although computation of sensor data quality has been explored in the literature, there hasn't been focus on how IoT applications can make use of such quality measures. In this section, we devise a quality model for evaluating sensor data. The model is composed of the following key IoT sensor data quality metrics:
\begin{enumerate}
    \item \emph{Suitability (S):} The sensor suitability estimates whether a sensor is suitable to meet the IoT application requirements.
    \item \emph{Accuracy (A):} The accuracy of sensor data is a measure of how accurately the sensor data represents the physical phenomenon it estimates. In this section, we show how sensor data accuracy may be estimated based on the stability of sensor data. However, we note that this may be replaced by other sensor accuracy measures existing in the literature.
    \item \emph{Completeness (C):} Sensor data completeness is an estimate of how much of the available sensor data was received by the IoT application within a time interval.
\end{enumerate}
These quality metrics are represented as a tuple \textit{(Q)}, ${Q} = \; <S, A, C>$. In this paper, we limit our discussion to the quality of sensor data, without considering the sensor hardware characteristics. However, we note that sensor hardware quality is crucial to IoT application performance and may be integrated as part of the IoT quality evaluation framework presented in the previous section. 

In the remainder of this section, we first define and measure each of the quality metrics introduced above. Subsequently, we aggregrate these metrics to calculate a quality value \textit{QV}, which provides a singular measure of the sensor data quality in the range $[0,1]$.
\subsection{Quality Metrics Tuple}

\subsubsection{Suitability}
The sensor suitability \textit{(S)} refers to whether a particular sensor is suitable for an application. If the application data requirements fall within the sensor measuring range, then a sensor is suitable for the application i.e. \textit{S = 1} else if the application data requirements fall out of the sensor measuring range making it unsuitable for the application, i.e. \textit{S = 0}. 
\begin{definition}
\textit{The sensor suitability defines whether the data value \textit{v}, expected from the application lies within the measuring interval of the sensor \textit{[a,b]} i.e. \textit{v} $\in$ \textit{[a,b]}}.
\end{definition}

To explain the suitability metric, let us consider an example of monitoring of milk temperature in milk cooling tanks, which is typically collected at a temperature within the range $30^\circ - 40^\circ C$ and subsequently needs to be cooled to $< 5^\circ C$. These
temperature values should be within the measuring range of the sensor $[a,b]$. The measuring range of a sensor is calibrated during its manufacturing. However, a temperature sensor that is suitable for the above application may not be suitable for another IoT application, such as one that detects the boiling of water (100$^\circ$ C), if \textit{v = 100$^\circ$ C} $\notin [a,b]$.

\subsubsection{Accuracy}
The sensor data accuracy represents the numerical precision of data collected from the sensor. 
The sensor may not always produce accurate data and the inaccuracy in data may occur due to various reasons such as interference from surrounding environment, malfunctioning of the sensor etc. In existing literature, sensor data accuracy is measured based on ground truth (GT), which is an alternate data source i.e. another sensor recording data values in the same application or an alternate data set i.e. historical data. Here, sensor data accuracy is measured based on the stability of sensor data, which makes it applicable to situations where there is no alternate data source to provide GT. \par

We compute the variation of data value \textit{v}, coming from a sensor, relative to its mean using moving standard deviation (mSD). As outlined, such a measure provides an estimate of the accuracy of sensor observations in terms of the stability over short time periods. This is because, high fluctuations in sensor data can lead to incorrect application outcomes. Nevertheless, we note that, alternative measures of sensor accuracy can be used where available, especially in the following cases: (a) when GT is available to provide a definite measure of sensor accuracy, and (b) when there are specific measures of accuracy for specific type of applications.
\begin{equation}
\label{eqn:msd}
    mSD_{k} = \sqrt{\frac{1}{m-1}\sum_{i=k-m+1}^k(v_i-\bar{v})^2)}, \: \forall k = 1,..., n
\end{equation}
where, \\
$v_i$ = data value at i\textsuperscript{th} row\\
$\bar{v}$ = average of data values \\
n = number of rows/data values\\
m = time window size defined by the application\\

The sensor data accuracy is quantified as standard error (SE) \cite{rel01}. Standard error is calculated as follows

\begin{equation}
\label{eqn:se}
    SE = \frac{mSD}{\sqrt{m}}
\end{equation}

Next, we compute the accuracy ($A$) based on mean of standard errors, i.e.,  $\overline{SE} = \frac {\sum\limits_{i=1}^k{SE}}{n/m}$ 

\begin{equation}
\label{eqn:acc_coefficient}
    A = (1-\overline{SE})
\end{equation} 

\begin{definition}
\textit{The accuracy \textit({$A$}) of a sensor data is defined as the precision in data measured in terms of moving standard deviation of a data value \textit{v} relative to its mean.} \end{definition}

\subsubsection{Completeness}
Completeness addresses the problem of missing sensor data values for a certain time window, which may be determined based on application requirements. According to the time delay between two consecutive recorded sensor data values, certain count of data values is expected for the defined time window, $count(Exp_v)$. If the sensor produces no value or NULL data values within the defined time window, it is marked as missing. Due to these missing values the actual count of recorded data ($count(Obt_v)$) values is less than $count(Exp_v)$. In our approach, we define completeness such that more is the number of missing values; less is the completeness. Mathematically, we compute data completeness as follows 
\begin{equation}
\label{eqn:comp}
   C = 1 - \frac{count(Exp_v) - count(Obt_v)} {count(Exp_v)}  
\end{equation}

The definition for data completeness is as follows 

\begin{definition}
\textit{The sensor data completeness refers to the degree to which sensor data values are not missing for a given time window.}
\end{definition}

For example, if a temperature sensor produces data values for every 1 millisecond and the application defines a time window  of 10 milliseconds, the sensor is expected to produce a total of 10 data values. Let us assume a sample of data values for 10 milliseconds as {16, 16.7, 16.5, \textit{NULL}, 16, 16.2, 16.8, 16.3, 16.1, \textit{NULL}}. In this sample data, the $count(Exp_v)$ is 10 and there are two missing values (as \textit{NULL}) and thus $count(Obt_v)$ is 8. Based on equation \ref{eqn:comp}, the completeness for this data set is 0.8.\par

\subsection{Computing Quality Value (QV) from the Quality Metrics Tuple }
In this subsection, we aggregate the quality metrics discussed in the above section to get a quality value \textit{(QV)} from the quality model. The input to the quality model is sensor data and application context information. The model measures these quality metrics based on the input it gets and computes  \textit{QV} value using the measurements received from suitability, accuracy and completeness. If a sensor is suitable for an application \textit{S=1}, and then the \textit{QV} value depends on the accuracy and completeness. Since, we assume that both accuracy and completeness affects \textit{QV} value equally, in equation (\ref{eqn:qvector}), we take average of accuracy and completeness values to compute \textit{QV}. When a sensor is not suitable (\textit{S=0}) then \textit{QV} will always evaluate to \textit{0}.


\begin{equation}
\label{eqn:qvector}
    QV = S \times \frac{(A + C)}{2}
\end{equation}

The quality model takes sensor data and application context information from the input module and measures suitability, accuracy and completeness. Based on the measures of these quality metrics, the quality model computes the quality value using equation (\ref{eqn:qvector}). We have used average to merge the measure of accuracy and completeness as it is the simplest statistical method to provide a composite measure. We have not chosen other methods like weighted average or RMS as the idea is only to provide a single value for the application quality. However, the proposed model can easily be extended to incorporate complex statistical methods.

The quality model provides the \textit{QV} to the SDQ\_Aware\_App module. The application logic uses the \textit{QV} to provide data quality aware applications SDQ\_Aware\_App. \par 

In the next section, we will use our proposed quality model and do a case study to demonstrate the applicability of the model.

\section{Case Study and Evaluation}

In this section, we present a case study of an IoT application and demonstrate how the quality model can be used by the application.

\subsection{Sample IoT Application}

We consider the scenario of monitoring the quality of milk storage in dairy farms. Dairy regulations typically require milk to be cooled to less than $5^{\circ} C$ within a few hours after milk is collected \cite{Rawmilkt32}. Failing to do so can result in the milk getting spoiled and having food safety risks.

An IoT application that uses temperature sensors to monitor the milk temperature can be used to provide real-time alerts to the farmer, so that appropriate action may be taken to avoid spoilage \cite{akbar2020iot}. However, sensor deployment in a farm environment is subject to a variety of factors which can affect the availability and even accuracy of the data. This makes it imperative for an IoT application to adapt to the sensor data quality, so as to deliver appropriate alerts.


Next, we have created an experimental setup to demonstrate the scenario of milk storage through an artificially simulated environment (the setup was placed inside a freezer).


\subsection{Experimental Setup}

In this section, we describe the test-bed for real-world evaluation of quality of sensor data at sensing stage of an IoT application. The proposed solution consists of an LM35 temperature sensor, an Arduino UNO board, three jumper wires and a laptop connected to the internet. LM35 temperature sensor is used to collect temperature data.\par
The proposed IoT testbed consists of an LM35 temperature sensor connected to an Arduino board using jumper wires and breadboard. 
Due to heterogeneity and interoperability in IoT, sensors are owned by different stakeholders and they are free to use the sensors of their choice. Considering this fact, our model is independent of the hardware used.  
The +5v for LM35 can be taken from the +5v out pin of Arduino UNO. The ground pin of LM35 is connected to GND pin of Arduino UNO and Vout, the analog out of LM35 is connected to A1 analog input pin of Arduino UNO.\par 

Arduino board which is connected to the laptop via USB cable, records the temperature data values. This temperature data is send to ThingSpeak which is an open source IoT platform. A python code is written to read the serial port and send the data to ThingSpeak where the data is collected in private channel. Besides collecting the data in real-time, ThingSpeak also supports data analysis through a well known numerical computing software, Matlab. ThingSpeak has integrated support from the Matlab software allowing its users to analyze and visualize the data uploaded on its channels. \par

The objective of the experiment is to evaluate the quality of data from a single source. For example, in our experiments we evaluate quality of data from LM35 temperature sensor. The experiment is conducted in an artificially simulated environment (the experimental setup was placed inside a freezer). The experiment run for 3 hours and 15 minutes and  612 temperature data values are collected, visualized and analysed during this duration on ThingSpeak. The collected temperature data values lie in the range between $28$ to $< 5^\circ C$. This is because in the artificial environment, the cooling is increased gradually from room temperature value to 0 degree. Some data values fall above 28 degree Celsius and some below 0 degree Celsius. These scattered data values act as outliers and are deviated from the expected temperature data values. 

In the following, we describe each module of the quality model for milk storage monitoring application.


\subsection{Evaluation of Quality Model based on Case Study}
In this subsection, we evaluate our proposed quality model and calculate $QV$ for the data collected from LM35 temperature sensor. We create an artificial environment for data collection where the temperature is brought down gradually from room temperature (\textasciitilde{28}°C) to  $< 5^\circ C$ which took 3 hours and 15 minutes. Next, based on our quality model, we compute the value of all the data quality metrics and aggregate them to provide a single quantitative measure for the application quality, i.e. $QV$. \par

The application designer selects the time window for data collection as per the application requirements and sensor measuring range from the sensor data sheet. The input to the quality model are as follows.
\begin{itemize}
   \item Sensor data : Temperature data values from LM35 temperature sensor.
    \item Context Information : Milking time, sensor measuring range, expected sensor data values per unit time, observed sensor data values per unit time.
\end{itemize}

 \label{fig:sdq}

Next, we describe how each of the metrics introduced previously can be incorporated in the milk storage monitoring application.
\subsubsection{Suitability of LM35 temperature sensor for milk storage monitoring application}
The quality model takes information about measuring range of LM35 temperature sensor and desired temperature for milk storage from the application context information. The measuring range of LM35 sensor is -55°C to 150°C. In milk storage monitoring application, the IoT application requirement is to achieve milk cooling to less than 5°C within $3.5$ hours after milk collection. Further, the model also checks about the LM35 sensor suitability for the milk storage monitoring application based on the suitability metric described in section IV(A). It returns \textit{S} with value assigned as \textit{S=1}. 

\subsubsection{Accuracy of LM35 temperature sensor data}
The model takes sensor data as input and  calculates its accuracy \textit{$A$} using equation \ref{eqn:msd} to \ref{eqn:acc_coefficient} in section IV(A). The total number of temperature data values collected \textit{N} is 612. We assume the value of time window size \textit{m} as 20. The value of $A$ is scaled down to make it always fall in the range between 0 and 1. 
The quality model based on the formula in equation \ref{eqn:se} and \ref{eqn:acc_coefficient}, computes \textit{SE} and \textit{A} as 0.172 and 0.828 respectively. 

\subsubsection{Completeness of LM35 temperature sensor data}
To measure the data completeness, the quality model takes expected and observed data values per unit time and computes the value for completeness \textit{C}, using eq. (\ref{eqn:comp}) of section IV(A). We program the arduino UNO to produce a data value for every 15 seconds. The experiment run for 11,700 seconds (3 hours and 15 minutes). The completeness module of the application model takes $count(Exp_v)$ = 780 while we observe the value of $count(Exp_v)$ = 612. This is due to the missing values in the data. It can be either no data received or NULL values in the collected data. Based on the formula in eq. (\ref{eqn:comp}) of section IV(A), the completeness module of the application model calculates the value of \textit{C} as 0.7846. \par

Now, based on the formula we proposed for evaluating the value of $QV$ in equation \ref{eqn:qvector}, the quality model  calculates $QV$ for LM35 sensor data using values obtained from  above measurements for the quality metrics as follows\\
$S=1$, $A=0.8280$  and $C=0.7846$ \\
$$QV = 1 \times \frac{(0.8280 + 0.7846)}{2}, using \; equation \; (\ref{eqn:qvector})$$
$$\Longrightarrow QV = 0.8063 $$\\

\begin{figure}[htbp]
  \centering
  \includegraphics[scale=0.35]{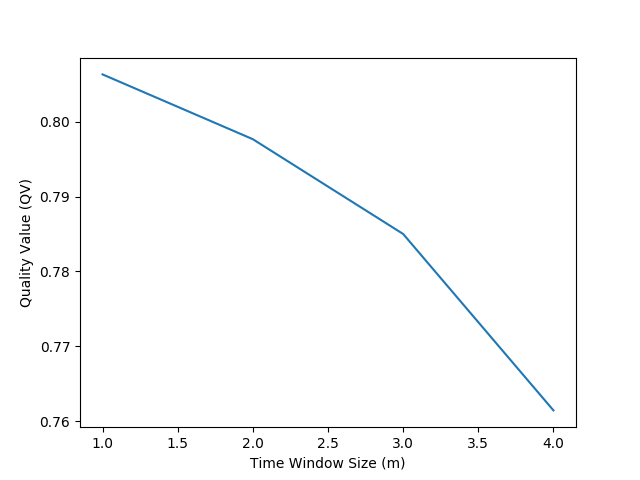}
  \caption{Effect of \textit{m} on \textit{QV}}
 \label{fig:m_qv}
\end{figure}

For the above QV, we have assumed that the application has defined the time window size (m) value as 20. However, we observed the effect of varying the value of \textit{m} on \textit{QV}. As shown in Fig. \ref{fig:m_qv}, initially we assume the value of \textit{m=50}, then in each iteration we double its value. We observe that as that when \textit{m} increases, the value of \textit{QV} decreases because on increasing \textit{m}, both \textit{A} and \textit{C} are impacted.  
The application uses $QV$ value to make a decision about the milk safety. In our application, we assume that if the value of $QV$ is equal or more than \textit{0.75} then the sensor data can be used to determine milk safety. 



\section{Conclusion and Future Directions}
In this paper, we proposed a framework for evaluation of IoT application quality. Subsequently, we presented a novel quality model to evaluate sensor data quality for Internet of Things applications. The quality model is based on input data, context information and IoT data quality metrics. We did a case study to evaluate the quality model based on sensor data quality, through which we highlight how the proposed metrics can be incorporated in the IoT application. 
Our future research directions are towards evaluating the quality of IoT data at all stages of an IoT application life cycle. Further, research in this area involves identifying and measuring the factors, apart from data, which can impact the application quality.

\bibliographystyle{IEEEtran}
\bibliography{sample-base}

\begin{thebibliography}{10}
\providecommand{\url}[1]{#1}
\csname url@samestyle\endcsname
\providecommand{\newblock}{\relax}
\providecommand{\bibinfo}[2]{#2}
\providecommand{\BIBentrySTDinterwordspacing}{\spaceskip=0pt\relax}
\providecommand{\BIBentryALTinterwordstretchfactor}{4}
\providecommand{\BIBentryALTinterwordspacing}{\spaceskip=\fontdimen2\font plus
\BIBentryALTinterwordstretchfactor\fontdimen3\font minus
  \fontdimen4\font\relax}
\providecommand{\BIBforeignlanguage}[2]{{%
\expandafter\ifx\csname l@#1\endcsname\relax
\typeout{** WARNING: IEEEtran.bst: No hyphenation pattern has been}%
\typeout{** loaded for the language `#1'. Using the pattern for}%
\typeout{** the default language instead.}%
\else
\language=\csname l@#1\endcsname
\fi
#2}}
\providecommand{\BIBdecl}{\relax}
\BIBdecl

\bibitem{yu2017recursive}
T.~Yu, X.~Wang, and A.~Shami, ``Recursive principal component analysis-based
  data outlier detection and sensor data aggregation in iot systems,''
  \emph{IEEE Internet of Things Journal}, vol.~4, no.~6, pp. 2207--2216, 2017.

\bibitem{maurer2006activity}
U.~Maurer, A.~Smailagic, D.~P. Siewiorek, and M.~Deisher, ``Activity
  recognition and monitoring using multiple sensors on different body
  positions,'' in \emph{International Workshop on Wearable and Implantable Body
  Sensor Networks (BSN'06)}.\hskip 1em plus 0.5em minus 0.4em\relax IEEE, 2006.

\bibitem{nesa2017iot}
N.~Nesa and I.~Banerjee, ``Iot-based sensor data fusion for occupancy sensing
  using dempster--shafer evidence theory for smart buildings,'' \emph{IEEE
  Internet of Things Journal}, vol.~4, no.~5, pp. 1563--1570, 2017.

\bibitem{ha2020sensing}
Q.~P. Ha, S.~Metia, and M.~D. Phung, ``Sensing data fusion for enhanced indoor
  air quality monitoring,'' \emph{IEEE Sensors Journal}, vol.~20, no.~8, pp.
  4430--4441, 2020.

\bibitem{banerjee2017iot}
T.~Banerjee and A.~Sheth, ``Iot quality control for data and application
  needs,'' \emph{IEEE Intelligent Systems}, vol.~32, no.~2, pp. 68--73, 2017.

\bibitem{tu2018data}
W.~Tu, ``Data-driven qos and qoe management in smart cities: A tutorial
  study,'' \emph{IEEE Communications Magazine}, vol.~56, no.~12, pp. 126--133,
  2018.

\bibitem{jayaraman2015yourself}
P.~P. Jayaraman, D.~Palmer, A.~Zaslavsky, and D.~Georgakopoulos,
  ``Do-it-yourself digital agriculture applications with semantically enhanced
  iot platform,'' in \emph{2015 IEEE tenth international conference on
  intelligent sensors, sensor networks and information processing
  (ISSNIP)}.\hskip 1em plus 0.5em minus 0.4em\relax IEEE, 2015, pp. 1--6.

\bibitem{jayaraman2015addressing}
P.~P. Jayaraman, D.~Palmer, A.~Zaslavsky, A.~Salehi, and D.~Georgakopoulos,
  ``Addressing information processing needs of digital agriculture with openiot
  platform,'' in \emph{Interoperability and Open-Source Solutions for the
  Internet of Things}.\hskip 1em plus 0.5em minus 0.4em\relax Springer, 2015,
  pp. 137--152.

\bibitem{hossain2011modeling}
M.~A. Hossain, P.~K. Atrey, and A.~E. Saddik, ``Modeling and assessing quality
  of information in multisensor multimedia monitoring systems,'' \emph{ACM
  Transactions on Multimedia Computing, Communications, and Applications
  (TOMM)}, vol.~7, no.~1, pp. 1--30, 2011.

\bibitem{kuemper2018valid}
D.~Kuemper, T.~Iggena, R.~Toenjes, and E.~Pulvermueller, ``Valid. iot: a
  framework for sensor data quality analysis and interpolation,'' in \emph{9th
  ACM Multimedia Systems Conference}, 2018, pp. 294--303.

\bibitem{rel01}
\BIBentryALTinterwordspacing
F.~T. L.~L. Jason L.~Huang, ``Standard error of measurement,'' 2016. [Online].
  Available:
  \url{https://www.britannica.com/science/standard-error-of-measurement}
\BIBentrySTDinterwordspacing

\bibitem{Rawmilkt32}
``Raw milk temperatures - department of agriculture,''
  \url{https://www.agriculture.gov.au/export/controlled-goods/dairy/registered-establishment/raw-milk-temperatures},
  November 2019, (Accessed on 04/27/2021).

\bibitem{akbar2020iot}
M.~O. Akbar, M.~J. Ali, A.~Hussain, G.~Qaiser, M.~Pasha, U.~Pasha, M.~S.
  Missen, N.~Akhtar \emph{et~al.}, ``Iot for development of smart dairy
  farming,'' \emph{Journal of Food Quality}, vol. 2020, 2020.

\end{thebibliography}


\end{document}